\documentclass[11pt,a4paper]{article} 
\pdfoutput=1
\usepackage{jheppub}

\usepackage{amsmath, amssymb}
\usepackage{concmath, palatino}
\usepackage{mathrsfs}
\usepackage{array,arydshln}

\usepackage{graphicx,epsfig}
\usepackage{epic}
\usepackage{color}

\newcommand{\be}{\begin{equation}}
\newcommand{\ee}{\end{equation}}
\newcommand{\ba}{\begin{eqnarray}}
\newcommand{\ea}{\end{eqnarray}}

\newcommand{\mc}{\mathcal }

\newcommand{\mk}{\mathfrak}
\newcommand{\hyper}[3]{{}_{3}F_{2}\left(\left.\begin{array}{c} #1 \\ #2 \end{array}\right | #3\right)}

\def\XXint#1#2#3{{\setbox0=\hbox{$#1{#2#3}{\int}$}
     \vcenter{\hbox{$#2#3$}}\kern-.5\wd0}}

    \newcommand{\beq}{\begin{equation}}
    \newcommand{\eeq}{\end{equation}}
    \newcommand\beqa{\begin{eqnarray}}
    \newcommand\eeqa{\end{eqnarray}}

\title{On the partition functions of higher spin black holes}

\author[a]{Matteo Beccaria } 
\author[a, b]{, Guido Macorini} 

\affiliation[a]{Dipartimento di Matematica e Fisica ìEnnio De Giorgi,\\
Universit\`a del Salento \& INFN, Via Arnesano, 73100 Lecce, 
Italy} 
                     
\affiliation[b]{Niels Bohr International Academy and Discovery Center,  \\
			Niels Bohr Institute, \\
		Blegdamsvej 17 DK-2100 Copenhagen, Denmark}

\emailAdd{matteo.beccaria@le.infn.it}
\emailAdd{macorini@nbi.ku.dk}

\abstract{
We reconsider black hole solutions of D = 3 higher-spin gravity in the $\mk{hs}[\lambda]\oplus \mk{hs}[\lambda]$
Chern-Simons formulation. A suitable generalisation of the BTZ black hole has a spin-3 chemical potential $\alpha$, 
and non-zero values of all the conserved charges associated with the asymptotic $\mc W_{\infty}[\lambda]$  symmetry. 
We  extend the available perturbative expansion of the partition function to order $\mc O(\alpha^{18})$ 
for generic values of the $\lambda$ parameter. The result matches the CFT prediction at 
$\lambda=0$ and at $\lambda=1$  where we provide the exact all-order expansion of the partition function.
The perturbative series is then analysed in the 
interesting non-trivial limit $\lambda\to \infty$ and we derive the exact analytical expressions of the partition function
and the spin-4 charge in closed form as functions of $\alpha$. Also, the first subleading correction at 
large $\lambda$ is shown to be 
simply related to the leading contribution.
}

\allowdisplaybreaks

\begin{document} \maketitle

\bigskip

\section{Introduction}

In the remarkable paper \cite{Gaberdiel:2010pz}, Vasiliev higher spin theory 
on $AdS_{3}$ \cite{Vasiliev:1995dn,Vasiliev:1996hn}  has been proposed to be 
holographically dual to certain coset minimal model conformal theories in the large-$N$ limit. In the simplest case, one considers
a consistent truncation where matter decouples and the pure higher spin theory  reduces to a Chern-Simons theory 
with algebra $\mk{hs}[\lambda]\oplus\mk{hs}[\lambda]$. Accordingly, the dual CFT has $W_{\infty}[\lambda]$  symmetry. 

\vskip 5pt
An important feature of these Chern-Simons theories  is that they admit  black hole solutions 
with non zero  values of the higher spin charges \cite{Gutperle:2011kf,Kraus:2011ds,Ammon:2012wc}.  In the holomorphic formalism \cite{Gutperle:2011kf,Ammon:2011nk}, it is possible to define 
a sensible thermodynamics and evaluate the black hole entropy from its partition function. The various charges 
are obtained by requiring that the gauge fields holonomy around a thermal circle is trivial. This condition 
fixes the relation between the charges and their conjugate chemical potentials. The entropy is then evaluated by 
the first law of thermodynamics. Recent developments based on this approach can be 
found in \cite{Ammon:2011ua,Kraus:2012uf,Chen:2012ba,David:2012iu,Kraus:2013esi,Gaberdiel:2013jca}. For a 
discussion of the relation between the holomorphic entropy and the canonical one, see comments at the end of 
section (\ref{sec:main-result}).

In the holomorphic formalism, higher spin black holes are interpreted as states of a dual 
CFT deformed by an irrelevant operator. These microscopic states can be counted by standard methods that exploit
the underlying $\mc W$ symmetry of the conformal theory. Agreement with the bulk calculation is found at relatively low
order in perturbation theory \cite{Gaberdiel:2011zw,Gaberdiel:2012yb,Gaberdiel:2013jca}. In particular, the $\mc O(\alpha^{8})$ expansion of the partition function in powers of the spin-3 charge 
chemical potential \cite{Kraus:2011ds} has been confirmed.

In this paper, we reconsider the Kraus-Perlmutter  calculation and extend it to the order $\mc O(\alpha^{18})$. 
This task does not involve any additional
conceptual tool and is a matter of computational techniques. Nevertheless, the result is, in our opinion, interesting for various reasons. The availability of a moderately long perturbation series allows to explore regularities and permits to 
propose conjectures about higher order effects. 

In particular, in Sec.~(\ref{sec:BHreview}) we review the setup of the Kraus-Perlmutter  calculation.
In Sec.~(\ref{sec:extended}), we present our results for the partition function at order $\mc O(\alpha^{18})$
and compare it with the CFT prediction at $\lambda=0,1$. At $\lambda=1$, we provide novel results for the 
all order CFT partition function. Finally, in Sec.~(\ref{sec:largelambda}), we consider the $\lambda\to\infty$ limit
of the $\mk{hs}[\lambda]$ algebra and report exact expressions for the partition function and the spin-4 charge.

\section{Black holes in higher spin gravity}
\label{sec:BHreview}

In this section, we briefly summarise some basic information about higher spin black holes in Chern-Simons gravity.
We start from the BTZ solution \cite{Banados:1992wn,Banados:1992gq} and present its extension 
with non zero spin charges according to the Kraus-Perlmutter construction \cite{Kraus:2011ds}. A nice review of the subject with much more details
can be found in \cite{Ammon:2012wc}.

\subsection{Chern-Simons gravity}

Einstein gravity with a negative cosmological constant can be written as a $SL(2,\mathbb R)\times SL(2,\mathbb R)$ Chern-Simons theory \cite{Achucarro:1987vz,Witten:1988hc}. The action involves the 
1-forms $(A, \overline A)$ taking values in the Lie algebra  $\mathfrak{sl}(2,\mathbb R)$
and reads
\be
S = S_{\rm CS}(A)-S_{\rm CS}(\overline A),
\ee
with
\be
S_{\rm CS} = \frac{k}{4\,\pi}\int\mbox{Tr}\bigg(A\wedge dA+\frac{2}{3}\,A\wedge A\wedge A\bigg).
\ee
The Chern-Simons level $k$ is related to the Newton constant $G_{\rm N}$ and the $AdS_{3}$ radius $\ell_{\rm AdS}$ by the
relation 
$k = \ell_{\rm AdS}/(4G_{\rm N})$.
When the Chern-Simons forms take value in a different Lie algebra $\mk g$, the resulting theory describes
Einstein gravity coupled to a set of
higher spin fields. An important example is $\mk g = \mk{sl}(N, \mathbb R)$ with integer $N\ge 3$. In this case,
one has the graviton and a tower of symmetric tensor fields of spin $s = 3,4,...,N$. In this paper, we shall focus on the
case of the infinite dimensional higher spin algebra $\mk{hs}[\lambda]$ where $\lambda$ is a positive real parameter 
determining the gravitational couplings among the various 
fields \cite{Prokushkin:1998bq,Campoleoni:2010zq}~\footnote{
The physical meaning of $\lambda$ is that it parameterizes a family of inequivalent $AdS$ vacua.
When scalar matter is coupled to gravity, it also sets the mass of the scalar fields, see for instance \cite{Ammon:2011ua}.}.
We also remark that the computation in $\mk{hs}[\lambda]$ with integer $\lambda=N$ 
is expected to reproduce the results in $\mk{sl}(N, \mathbb R)$.

\subsection{The BTZ black hole}

The $SL(3,\mathbb R) \times SL(3,\mathbb R)$  Chern-Simons theory admits 
smooth, asymptotically $AdS_{3}$, 
black hole solutions with a spin-3 charge and sensible thermodynamics \cite{Gutperle:2011kf,Ammon:2011nk}.
The original BTZ black hole \cite{Banados:1992wn,Banados:1992gq} does not involve the spin-3 charge 
and we very briefly review its construction as a preparation to its higher spin generalisation.
Denoting the generators of a $\mk{sl}(2, \mathbb R)$ subalgebra  as $\{L_{-1}, L_{0}, L_{1}\}$, 
the BTZ solution reads
\ba
A &=& \bigg(e^{\rho}L_{1}-\frac{2\pi}{k}\,\mc L\,e^{-\rho}\,L_{-1}\bigg)\,dx^{+}+L_{0}\,d\rho, \\
\overline A &=& -\bigg(e^{\rho}L_{-1}-\frac{2\pi}{k}\,\overline{\mc L}\,e^{-\rho}\,L_{1}\bigg)\,dx^{-}-L_{0}\,d\rho, 
\ea
where $(\rho, x^{\pm}\equiv t\pm \varphi)$ are the space-time coordinates and $\mc L, \overline{\mc L}$
are linear combinations of the conserved mass and angular momentum charges.

The Euclidean BTZ black hole is obtained by taking $dx^{+} = dz$ and $dx^{-} = −d\overline z$ and 
making the identification $(z, \overline z)\sim (z+2\pi \tau, \overline z+2\pi \overline \tau)$
in order to avoid a conical singularity at the horizon. Focusing on holomorphic quantities, the conserved charge 
is 
\be
\mc L = -\frac{k}{8\,\pi\,\tau^{2}},
\ee
where the parameter $\tau$ is a function of the inverse temperature and horizon angular velocity of the black hole.
This relation follows from the requirement of a smooth horizon.
It is important to have a gauge invariant parametrisation of the BTZ black hole. This is achieved by introducing
the holonomy of the gauge connection around the Euclidean time circle defined above. This is 
\be
\omega = 2\,\pi\,(\tau A_{+}-\overline \tau A_{-}),
\ee
whose eigenvalues $\{0, \pm 2\,\pi\,i\}$ are encoded in the trace relations
\be
\label{eq:BTZ-trace}
\mbox{Tr}(\omega) = 0, \quad \mbox{Tr}(\omega^{2}) = -8\,\pi^{2}, \quad \mbox{Tr}(\omega^{3}) = 0.
\ee
The inclusion of a spin-3 charge $\mc W$ (and its anti-holomorphic counterpart) requires the introduction
of an associated chemical potential $\mu$. Again, smoothness of the horizon is expected to fix the full dependence
of $\mc L$ and $\mc W$ on $\tau$ and the chemical potential. As discussed in \cite{Gutperle:2011kf,Ammon:2011nk}, smoothness
has to be enforced in a $SL(3,\mathbb R)\times SL(3,\mathbb R)$ gauge invariant way by imposing 
the holonomy trace conditions (\ref{eq:BTZ-trace}).
Also, this BTZ holonomy prescription guarantees a sensible thermodynamics.  The black hole must be the saddle point
contribution to a partition function of the form (we write only the holomorphic part)
\be
Z(\tau, \alpha) = \mbox{Tr}\bigg[e^{4\pi^{2}i(\tau \mc L+\alpha \mc W)}\bigg].
\ee
Hence, the following integrability condition must hold: $\partial_{\alpha}\mc L=\partial_{\tau}\mc W
\sim \partial^{2}_{\alpha\tau}Z$. This
leads to a precise relation $\alpha=\overline\tau\,\mu$ between $\alpha, \tau$, and $\mu$ and to a well defined entropy in agreement with the first law
of thermodynamics.

\subsection{The Kraus-Perlmutter construction}

The Kraus-Perlmutter construction \cite{Kraus:2011ds} leads to 
a generalisation of the BTZ black hole suitable for higher spin 
gravity based on the algebra $\mk{hs}[\lambda]\oplus\mk{hs}[\lambda]$ (see Appendix A for the details relevant to this paper). The main ingredient is  the 
connection $a_{+}$ which is related to $A$ by a gauge transformation \cite{Campoleoni:2010zq}\footnote{
This gauge transformation is such that we work with a flat connection without $\rho$ dependence and with no $\rho$ component~\cite{Coussaert:1995zp}.}
\be
a_{+} = V^{2}_{1}-\frac{2\,\pi\,\mc L}{k}\,V^{2}_{-1}-\mc N(\lambda)\,\frac{\pi\,\mc W}{2\,k}\,V^{3}_{-2}+
\mc J_{4}\,V^{4}_{-3}+\mc J_{5}\,V^{5}_{-4}+\cdots, 
\ee
where $V^{s}_{m}$ are $\mk{hs}[\lambda]$ generators.
The normalisation $\mc N(\lambda) = \sqrt\frac{20}{\lambda^{2}-4}$ is chosen to simplify comparison 
to the $SL(3,\mathbb R)$ results of \cite{Gutperle:2011kf,Ammon:2011nk} and is the same as in  \cite{Kraus:2011ds}. The connection $a_{+}$ contains
the Einstein gravity charge $\mc L$, the spin 3 charge $\mc W$ and additional charges $\mc J_{n\ge 4}$ for all higher spin 
fields. The solution to the holonomy equations will have to satisfy the integrability condition $\mc L_{\alpha}=\mc W_{\tau}$. The holonomy is 
\be
\omega = 2\,\pi\,\bigg[\tau\,a_{+}-\alpha\,\mc N(\lambda)\,\bigg(a_{+}\star a_{+}-\frac{2\,\pi\,\mc L}{3k}
(\lambda^{2}-1)\bigg)\bigg],
\ee
and the smoothness condition reads simply
\be
\label{eq:smooth}
\mbox{Tr}(\omega^{n}) = \mbox{Tr}(\omega_{\rm BTZ}^{n}) = \frac{24}{\lambda\,(\lambda^{2}-1)}\,
\lim_{t\to 0}\left(\partial_{t}^{n}\frac{\sin(\pi\,\lambda\,t)}{\sin(\pi\,t)}\right),
\ee
where
\be
\omega_{\rm BTZ} = 2\,\pi\,\tau\,\bigg(V^{2}_{1}+\frac{1}{4\,\tau^{2}}\,V^{2}_{-1}\bigg),
\ee
and the r.h.s of  (\ref{eq:smooth}) has been derived in \cite{Gaberdiel:2013jca} by exploiting the fact that the $V^{s}_{m}$ generators in $\omega_{\rm BTZ}$ have all $s=2$.

\section{Extended perturbative analysis}
\label{sec:extended}

We have solved the holonomy conditions (\ref{eq:smooth}) perturbatively in $\alpha$ by extending the analysis of \cite{Kraus:2011ds}
in order to compute the partition function up to terms $\mc O(\alpha^{18})$ . The computational complexity
is rather high and the main difficulty is in the evaluation of the traces $\mbox{Tr}(\omega^{n})$. To this aim, we have 
split 
\be
\omega = \widetilde\omega+ \mc N(\lambda)\,\frac{4\pi^{2}\alpha\,\mc L}{3k}
(\lambda^{2}-1),
\ee
and computed the traces $\mbox{Tr}(\widetilde\omega^{n})$. The holonomy $\widetilde \omega$ can be
separated out according to the $m$ value of the various $V^{s}_{m}$ generators
\be
\widetilde \omega=\sum_{m=-\infty}^{2}\widetilde \omega_{m}.
\ee
Each non vanishing contribution to  $\mbox{Tr}(\widetilde\omega^{n})$ comes from  terms of the form 
\be
\widetilde\omega_{m_{1}}\star\cdots\star \widetilde\omega_{m_{n}},\qquad m_{1}+\cdots m_{n}=0.
\ee
Also, ciclicity invariance 
\be
\mbox{Tr}(\widetilde\omega_{m_{1}}\star\widetilde\omega_{m_{2}}\star\cdots\star \widetilde\omega_{m_{n}})=
\mbox{Tr}(\widetilde\omega_{m_{2}}\star\cdots\star \widetilde\omega_{m_{n}}\star\widetilde\omega_{m_{1}}),
\ee
is used to greatly reduce the number of relevant traces. Once, the holonomy conditions (\ref{eq:smooth}) have been
generated, they can be solved by the Ansatz
\be
\label{eq:LW-exp}
\mc L = k\,\sum_{p=0}^{\infty} \mc L^{(p)}(\lambda)\,\frac{\alpha^{2p}}{\tau^{4p+2}}, \qquad
\mc W = k\,\sum_{p=0}^{\infty} \mc W^{(p)}(\lambda)\,\frac{\alpha^{2p+1}}{\tau^{4p+5}}, 
\ee
as well as
\be
\label{eq:J-exp}
\mc J_{n} = \sum_{p=0}^{\infty} \mc J_{n}^{(p)}(\lambda)\,\frac{\alpha^{n-2+2p}}{\tau^{3n+4p-4}},\qquad n\ge 4.
\ee

\subsection{The $\mc O(\alpha^{18})$ partition function}
\label{sec:main-result}

We can obtain the partition function by integrating the expansion of $\mc L$ according to  the relation
 \be
 \label{eq:Ldef}
 \partial_{\tau}\log Z=4\pi^{2}i\,\mc L.
 \ee
 Notice also that, remarkably, the expansion of $\mc W$ obeys the integrability constraint at the considered perturbative
 order. 
The general form of the partition function is then
\be
\log Z(\tau, \alpha)  = \sum_{p=0}^{\infty} Z^{(p)}(\lambda)\,\frac{\alpha^{2p}}{\tau^{4p+1}},
\ee
and its explicit value is found to be
\ba
\label{eq:mainresult}
\lefteqn{\log Z(\tau, \alpha) =} && \,\,\quad \qquad\qquad \frac{i\,\pi\,k}{2\,\tau}\bigg[
1-\frac{4}{3}\frac{\alpha^{2}}{\tau^{4}}+\frac{400}{27}\frac{\lambda ^2-7}{\lambda
   ^2-4}\frac{\alpha^{4}}{\tau^{8}}-\frac{1600}{27}\frac{5 \lambda ^4-85 \lambda ^2+377}{\left(\lambda
   ^2-4\right)^2}\frac{\alpha^{6}}{\tau^{12}}\nonumber\\
   &&+\frac{32000}{81}\frac{20 \lambda ^6-600 \lambda ^4+6387 \lambda
   ^2-23357}{ \left(\lambda ^2-4\right)^3}\frac{\alpha^{8}}{\tau^{16}}\nonumber \\
   &&
   -\frac{640000}{1701}\frac{665 \lambda
   ^8-30590 \lambda ^6+571494 \lambda ^4-4982450 \lambda ^2+16493303}{\left(\lambda
   ^2-4\right)^4}\frac{\alpha^{10}}{\tau^{20}}\nonumber \\
   &&
   +\frac{12800000}{729}\frac{506 \lambda ^{10}-32890 \lambda ^8+940073 \lambda
   ^6-14337947 \lambda ^4+112660375 \lambda ^2-351964697}{\left(\lambda
   ^2-4\right)^5}\frac{\alpha^{12}}{\tau^{24}}\nonumber \\
   && -\frac{51200000}{243 \left(\lambda ^2-4\right)^6} (1625 \lambda ^{12}-141375 \lambda ^{10}+5707245
   \lambda ^8-133009165 \lambda ^6+1824202749 \lambda ^4\nonumber \\
   && -13483461495 \lambda
   ^2+40593663941)\frac{\alpha^{14}}{\tau^{28}}\nonumber \\
   &&+\frac{1024000000}{45927 \left(\lambda ^2-4\right)^7} (629300
   \lambda ^{14}-70481600 \lambda ^{12}+3810410205 \lambda ^{10}-125413188775 \lambda
   ^8\nonumber \\
   && +2625581492023 \lambda ^6-33808153697493 \lambda ^4+240164027951297 \lambda
   ^2-704782787198207)\frac{\alpha^{16}}{\tau^{32}}\nonumber \\
   && -\frac{20480000000}{137781 \left(\lambda
   ^2-4\right)^8} (4057900 \lambda ^{16}-568106000 \lambda ^{14}+39583087600 \lambda ^{12}-1745724243200
   \lambda ^{10}\nonumber \\
   &&+51602653138939 \lambda ^8-1013197506287120 \lambda ^6+12516285865006996 \lambda
   ^4\nonumber \\
   &&-86465944436086340 \lambda ^2+249020069093788675)\,\frac{\alpha^{18}}{\tau^{36}}+\dots\bigg].
\ea
The expansion of $\mc J_{4}$ is also interesting and reads
\ba
\mc J_{4} &=&
\frac{35}{9} \frac{1}{\lambda ^2-4}\frac{\alpha^{2}}{\tau^{8}}-\frac{700}{9}\frac{2 \lambda
   ^2-21}{\left(\lambda ^2-4\right)^2}\frac{\alpha^{4}}{\tau^{12}}
   +\frac{2800}{9}\frac{20
   \lambda ^4-480 \lambda ^2+3189}{\left(\lambda ^2-4\right)^3}\frac{\alpha^{6}}{\tau^{16}}\nonumber\\
   &&-\frac{8000}{81}\frac{2660 \lambda ^6-107730 \lambda ^4+1626345 \lambda ^2-8871827}{
   \left(\lambda ^2-4\right)^4}\frac{\alpha^{8}}{\tau^{20}}\nonumber \\
   &&+\frac{1120000}{243}\frac{2530 \lambda
   ^8-151800 \lambda ^6+3855213 \lambda ^4-47653816 \lambda ^2+233903943}{\left(\lambda
   ^2-4\right)^5}\frac{\alpha^{10}}{\tau ^{24}}\nonumber \\
   &&-\frac{4480000}{27\,(\lambda^{2}-4)^{6}}\,(3250 \lambda ^{10}-268125 \lambda
   ^8+10073985 \lambda ^6-209221150 \lambda ^4\nonumber \\
   &&+2323193928 \lambda ^2-10693046847)
\frac{\alpha^{12}}{\tau ^{28}}+\dots .
\ea
At this point, we need to comment about the meaning of the entropy that can be derived from this partition
function. It is obtained in the so-called holomorphic approach that we have followed. 
Although the partition function matches the CFT result in \cite{Gaberdiel:2011zw}, it was shown in 
\cite{Perez:2013xi,deBoer:2013gz} that the associated entropy does not 
agree with the canonical entropy. The discrepancy is fully discussed in \cite{Compere:2013gja,Compere:2013nba}
 where it has been showed how to define 
thermodynamical variables so that the canonical partition function has a natural CFT interpretation. 
Remarkably, the results in the holomorphic formalism can be adapted to the canonical definition by simply
replacing $\tau$ and $\alpha$ by $\tilde{\tau}$ and $\tilde{\alpha}$
defined in section 4 of \cite{Compere:2013nba}.

\subsection{Comparison with CFT in the $\lambda=0,1$ limits}

\begin{figure}[htb]
\begin{center}
\includegraphics[scale=0.6]{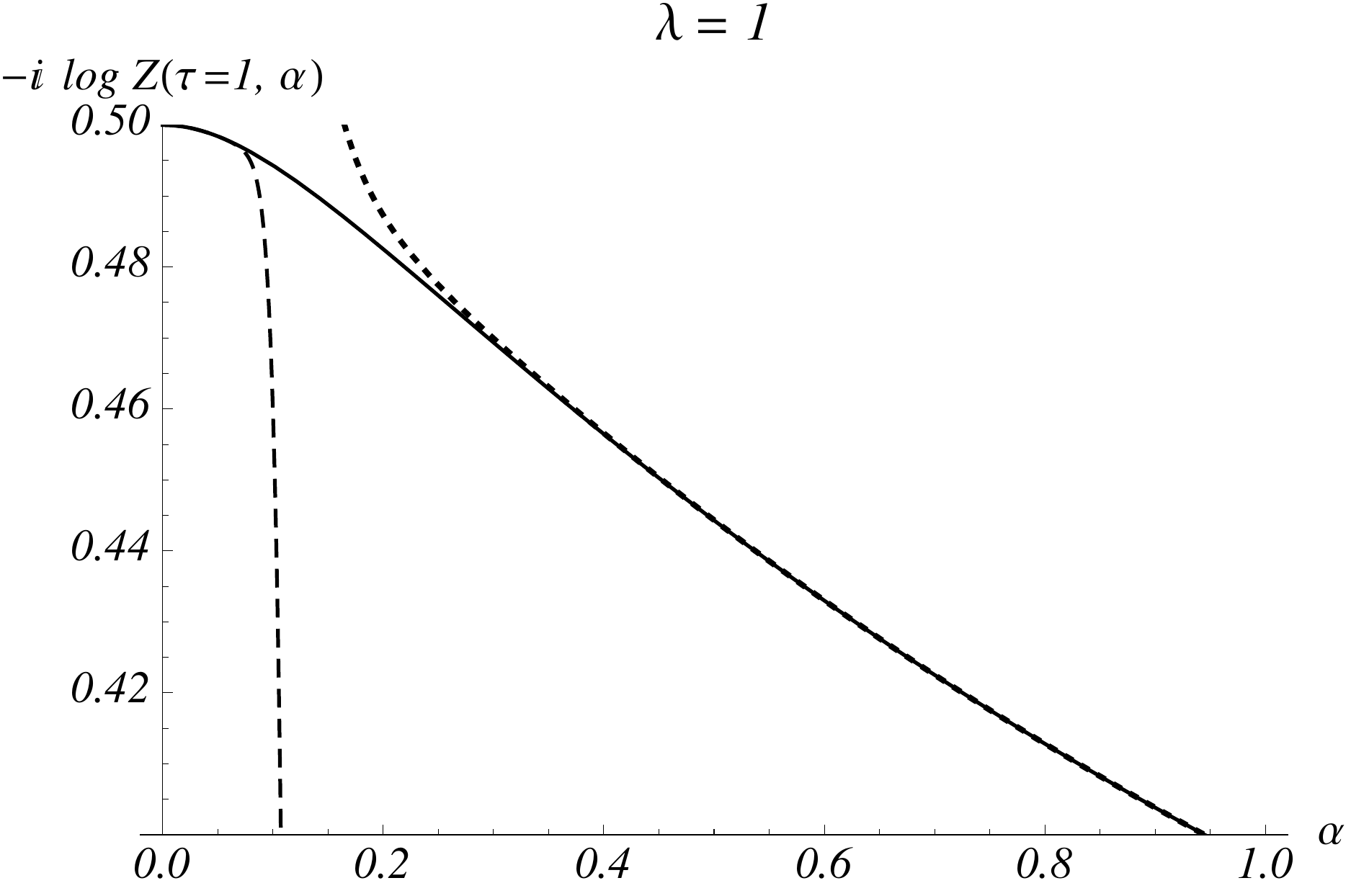}
\caption{Plot of $-i\,\log Z_{\lambda=1}$ at $\tau=1$. The dashed and dotted lines are the obtained by computing a few terms of the weak and strong coupling expansions.}
\label{fig:logZ1}
\end{center}
\end{figure}

The values at $\lambda=0,1$ can be checked against a simple CFT computation. For $\lambda=1$,
the $\mk{hs}[1]$ algebra admits a realisation in terms of free complex bosons (see \cite{Kraus:2011ds}
for the precise matching).
In App.~(B.1), we derive a new closed formula for the generic term of the perturbative expansion of the CFT 
partition function. The result is 
\be
\label{eq:exact1}
\log Z_{\lambda=1}(1,\alpha) = \frac{6\,\pi\,i}{\sqrt\pi}\,\sum_{n=0}^{\infty}
\frac{B_{2n+2}\,\Gamma\left(2n+\frac{1}{2}\right)}{(2n+2)!}\,\left(\frac{320}{3}\right)^{n}\,\alpha^{2n},
\ee
where $B_{2n}$ are Bernoulli numbers. 
Explicitly, this reads
\ba
\log Z_{\lambda=1}(1,\alpha) &=& i\,\pi\bigg(
\frac{1}{2}-\frac{2 \alpha ^2}{3}+\frac{400 \alpha ^4}{27}-\frac{8800 \alpha ^6}{9}+\frac{10400000
   \alpha ^8}{81}-\frac{142843520000 \alpha ^{10}}{5103}\nonumber \\
   &&+\frac{6656384000000 \alpha
   ^{12}}{729}-\frac{1011602560000000 \alpha ^{14}}{243} \\
   &&+\frac{116100858752000000000 \alpha
   ^{16}}{45927}-\frac{12939787022848000000000 \alpha ^{18}}{6561}\bigg)+\dots ,\nonumber
\ea
and is in full agreement with our generic $\lambda$ result when $\lambda$ is set to 1. Notice also that, as far as the 
small $\alpha$ expansion is concerned, we can write
\be
\log Z_{\lambda=1}(1,\alpha) = i\,\pi\bigg[
-\frac{3}{40\,\alpha^{2}}+\frac{3}{4}\,\sqrt\frac{3}{5}\,\frac{1}{\alpha}-3\,\sqrt{2}\,\sqrt[4]{\frac{3}{5}}
\,\zeta\left(-\frac{1}{2}, \frac{1}{8\,\alpha}\,\sqrt\frac{3}{5}\right)\,\frac{1}{\sqrt\alpha}
\bigg],
\ee
where the Hurwitz $\zeta$ function is $\zeta(a, s) = \sum_{n=0}^{\infty}\frac{1}{(a+n)^{s}}$.
This expression can be used to expand the partition function at large $\alpha$, at least assuming that no non-perturbative 
effects correct it~\footnote{See the discussion in the concluding section for comments concerning the convergence of the perturbative expansions presented
here and in the following}. One finds
\be
\log Z_{\lambda=1}(1,\alpha) = i\,\pi\bigg[
\frac{3 \sqrt[4]{\frac{3}{5}}  \zeta
   \left(\frac{3}{2}\right)}{2 \sqrt{2} \pi }\,\frac{1}{\sqrt\alpha}-\frac{3 \sqrt{\frac{3}{5}}}{4
   \alpha }-\frac{3
   \left(\frac{3}{5}\right)^{3/4} \zeta
   \left(\frac{1}{2}\right)}{8 \sqrt{2}}\,\frac{1}{\alpha^{3/2}}-\frac{3}{40 \alpha
   ^2}+\dots
\bigg].
\ee
A plot of $-i\,\log Z_{\lambda=1}$ as a function of $\alpha$ for $\tau=1$ is shown in Fig.~(\ref{fig:logZ1})
where we also superimpose the (asymptotic) weak and (convergent) strong coupling expansions

For $\lambda=0$, we can compare with a free theory of  complex fermions 
with symmetry $\mc W_{1+\infty}$, see again \cite{Kraus:2011ds}. This has spin content $s=1,2,3,\dots$ , and is related to $\mk{hs}[0]$
by a constraint that eliminates the spin-1 current. In App.~(B.2), we develop some technical results
needed to efficiently compute the $\mc O(\alpha)^{18}$ expansion of the CFT partition function. We  find again full agreement.

\subsection{Properties of the polynomials $Z^{(p)}(\lambda)$}

As remarked in \cite{Kraus:2011ds}, the zeroes of the polynomials $Z^{(p)}(\lambda)$ depend on $p$ and it is
difficult to assign a definite meaning to them.  Nevertheless, using our extended data, we observe that there is a root 
moving toward  $\lambda=3$, as $p$ is increased. The same seems to happen around $\lambda=4$
for large enough $p$.  This can be seen in the following table ( a dash means that there only one root near 3 and no other one near 4)
\be
\begin{array}{l|l}
p & {\rm root\ nearest\ to\ 3} \\
\hline
2 &2.64575131106 \\
3 &2.91547597523\\
4 &2.96149892228\\
5 &3.00462876498\\
6 &2.99974422033\\
7 &3.00001095940\\
8 &2.99999964017\\
9 &3.00000000939
\end{array}\qquad\qquad
\begin{array}{l|l}
p & {\rm root\ nearest\ to\ 4} \\
\hline
2 & - \\
3 & - \\
4 & - \\
5 & - \\
6 & - \\
7 & 3.90889802016\\
8 & 4.02822441758\\
9 & 3.99768663666
\end{array}
\ee
From this numerical analysis, it is tempting to conjecture that the polynomials $Z^{(p)}(\lambda)$ have roots 
converging to the integers $3, 4, \dots$ as $p\to \infty$. This property is consistent with the truncation of
the underlying $\mc W_{\infty}$ algebra that happens precisely at these values of $\lambda$. The
 values $Z^{(p)}(\lambda=N)$ decay relatively fast as $p\rightarrow\infty$ for $N=3,4,\dots$, and this is
suggested by the fact that there is a nearby zero.

\section{Exact results in the  $\lambda=\infty$ limit} 
\label{sec:largelambda}

An inspection of the explicit expansions (\ref{eq:LW-exp}) and (\ref{eq:J-exp}) shows that $\mc L$, $\mc W$,
and the scaled charges $\lambda^{n-2}\,\mc J_{n}$ admit a smooth non-trivial limit when $\lambda\to\infty$. 
We remark that the higher spin charges $\mc J_{n}$ are not negligible despite the fact that they scale like
$1/\lambda^{n-2}$. Indeed, they appear in the holonomy equations with coefficients involving powers of $\lambda$
and leaving a non zero contribution (see, for example, (\ref{eq:trace2}) in the following). In this section, we 
shall present results about this limit. In particular, we derive the closed expression for the partition function and
for the scaled spin-4 charge $\lambda^{2}\,\mc J_{4}$, both at leading and next-to-leading order at large $\lambda$.
More comments about the meaning of the large $\lambda$ limit are deferred to the concluding section. 

\subsection{Partition function}

\begin{figure}[htb]
\begin{center}
\includegraphics[scale=0.6]{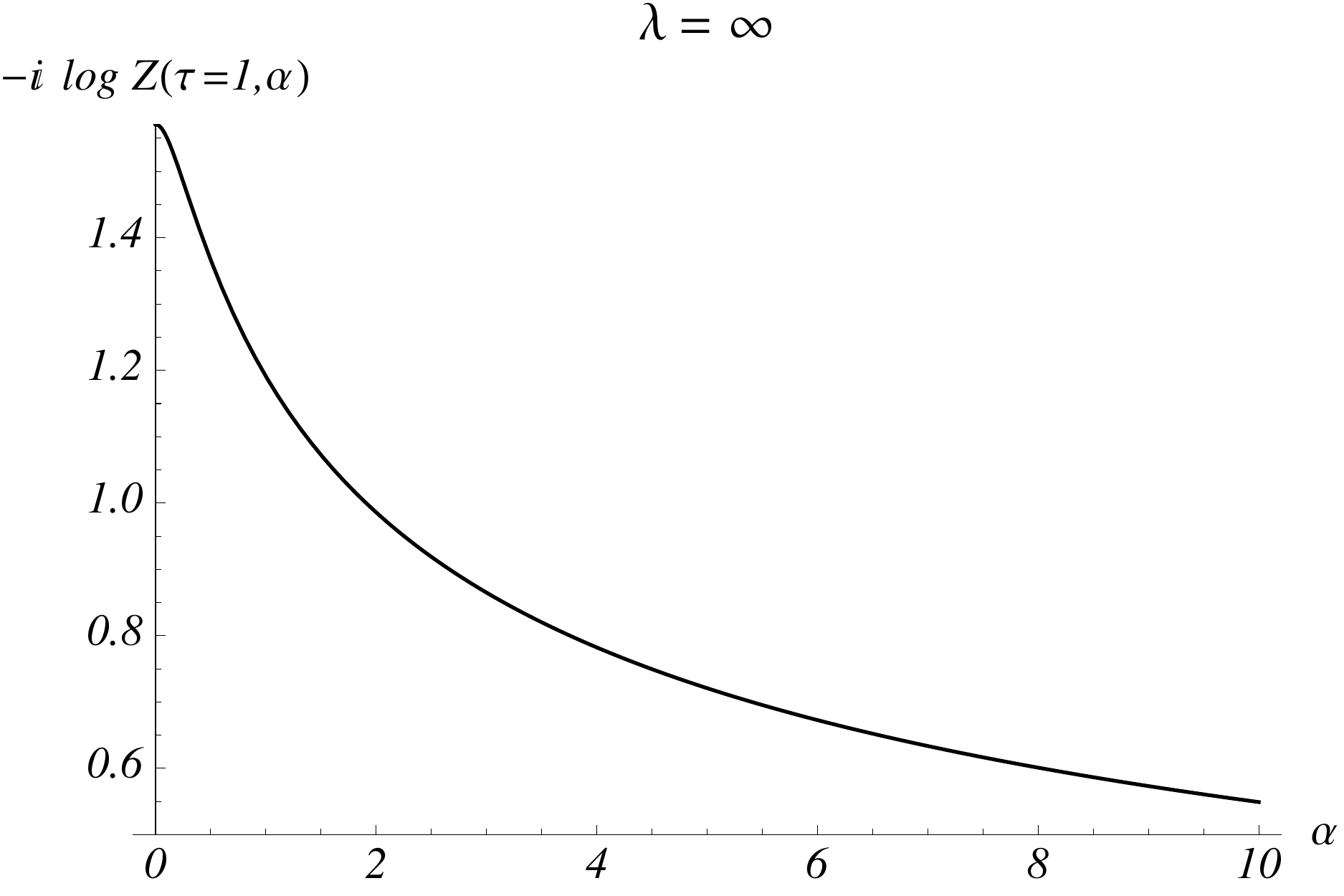}
\caption{Plot of $-i\,\log Z_{\lambda=\infty}$ at $\tau=1$}
\label{fig:logZ}
\end{center}
\end{figure}

The partition function admits a smooth limit for $\lambda\to\infty$. This limit is definitely non trivial since all 
higher spin charges are switched on. The series is explicitly
\ba
\lefteqn{\log Z_{\lambda=\infty}(\tau, \alpha) = }  && \quad \qquad\qquad\qquad \frac{i\,\pi\,k}{2\,\tau}\bigg(
1-\frac{4 \alpha ^2}{3 \tau ^4}+\frac{400 \alpha ^4}{27 \tau ^8}-\frac{8000 \alpha ^6}{27 \tau
   ^{12}}+\frac{640000 \alpha ^8}{81 \tau ^{16}}-\frac{60800000 \alpha ^{10}}{243 \tau
   ^{20}} \\
   &&+\frac{6476800000 \alpha ^{12}}{729 \tau ^{24}}-\frac{83200000000 \alpha ^{14}}{243 \tau
   ^{28}}+\frac{92057600000000 \alpha ^{16}}{6561 \tau ^{32}}-\frac{11872256000000000 \alpha
   ^{18}}{19683 \tau ^{36}}+\dots\bigg).\nonumber
\ea
Inspection of the coefficients suggests the following remarkable form for the generic term
\be
\log Z_{\lambda=\infty}(\tau, \alpha) = \frac{12\,i\,\pi\,k}{\tau}\,
\sum_{p=0}^{\infty}(-1)^{p}\,\left(\frac{20}{3}\right)^{p}\,\frac{\Gamma(4p)}{\Gamma(p)\Gamma(3p+4)}\,
\frac{\alpha^{2p}}{\tau^{4p}},
\ee
where the $p=0$ term has to be evaluated with $\Gamma(4p)/\Gamma(p)\to 1/4$. The infinite series can be summed
 and gives a closed expression~\footnote{Clearly, the hypergeometric nature of this expression means that it cannot be obtained by any finite truncation at $\lambda=N$ because in that case the holonomy constraints
are algebraic.}
\be
\log Z_{\lambda=\infty}(\tau, \alpha) = \frac{3\,i\,\pi\,\tau^{3}}{160\,\alpha^{2}}\,\bigg[
\hyper{-\frac{3}{4}, \ -\frac{1}{2}, \ -\frac{1}{4}}{\frac{1}{3}, \ \frac{2}{3}}{-\frac{5120\,\alpha^{2}}{81\,\tau^{4}}}
-1
\bigg].
\ee
From this expression, we can expand at large $\alpha/\tau^{2}$ with the result
\ba
\log Z_{\lambda=\infty}(\tau, \alpha) &=&
 -\frac{i \pi \Gamma \left(-\frac{3}{4}\right) \Gamma
   \left(\frac{1}{4}\right)}{12 \sqrt{3} \sqrt[4]{5} \Gamma \left(\frac{13}{12}\right) \Gamma
   \left(\frac{17}{12}\right)}\,\frac{1}{\alpha^{1/2}}
   -\frac{i \sqrt{\frac{3}{5}} \pi}{2}\,\frac{\tau}{\alpha}
   +\frac{i \sqrt{3}
   \pi   \Gamma \left(-\frac{1}{4}\right)^2}{64\ 5^{3/4}
   \Gamma \left(\frac{7}{12}\right) \Gamma \left(\frac{11}{12}\right)}\,\frac{\tau^{2}}{\alpha^{3/2}}
   \nonumber \\
&&   -\frac{3 i \pi  }{160}
\,\frac{\tau^{3}}{\alpha^{2}}-\frac{i \sqrt{3} \pi   \Gamma
   \left(-\frac{3}{4}\right) \Gamma \left(\frac{1}{4}\right)}{32768 \sqrt[4]{5} \Gamma
   \left(\frac{13}{12}\right) \Gamma \left(\frac{17}{12}\right)}\,\frac{\tau^{4}}{\alpha^{5/2}}+\frac{3 i \sqrt{\frac{3}{5}} \pi 
   }{25600}\,\frac{\tau^{5}}{\alpha^{3}}+\dots.
   \ea
A plot of $-i\,\log Z$ as a function of $\alpha$ for $\tau=1$ is shown in Fig.~(\ref{fig:logZ}).

\begin{figure}[htb]
\begin{center}
\includegraphics[scale=0.6]{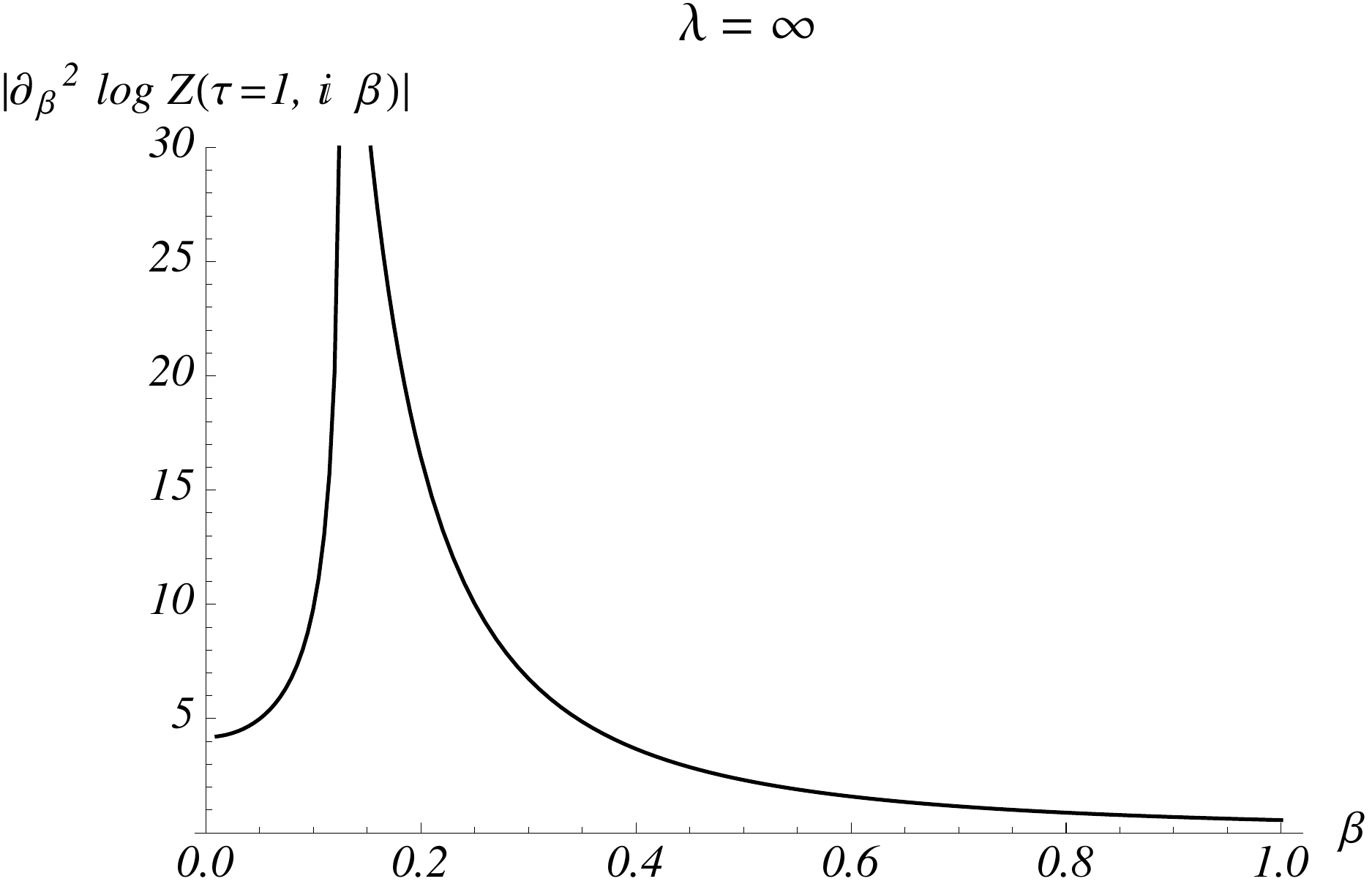}
\caption{Plot of $|\partial^{2}_{\beta} \log Z(\tau=1, i\, \beta)|$ at $\tau=1$. The singular point
is located at (\ref{eq:sing}).}
\label{fig:Chi}
\end{center}
\end{figure}

The curve is smooth for all real values of $\alpha/\tau^{2}$. Nevertheless, the hypergeometric function
has a branch point at the imaginary points
\be
\label{eq:sing}
\left(\frac{\alpha}{\tau^{2}}\right)_{\rm branch} = \pm \frac{9}{32\,\sqrt 5}\,i\, \simeq \pm 0.126\,i.
\ee
In Fig.~(\ref{fig:Chi}), we plot the susceptibility-like quantity
$|\partial^{2}_{\beta} \log Z(\tau=1, i\, \beta)|$ in order to show the effect of the branch point.

\subsection{Spin-4 charge $\mc J_{4}$}

\begin{figure}[htb]
\begin{center}
\includegraphics[scale=0.6]{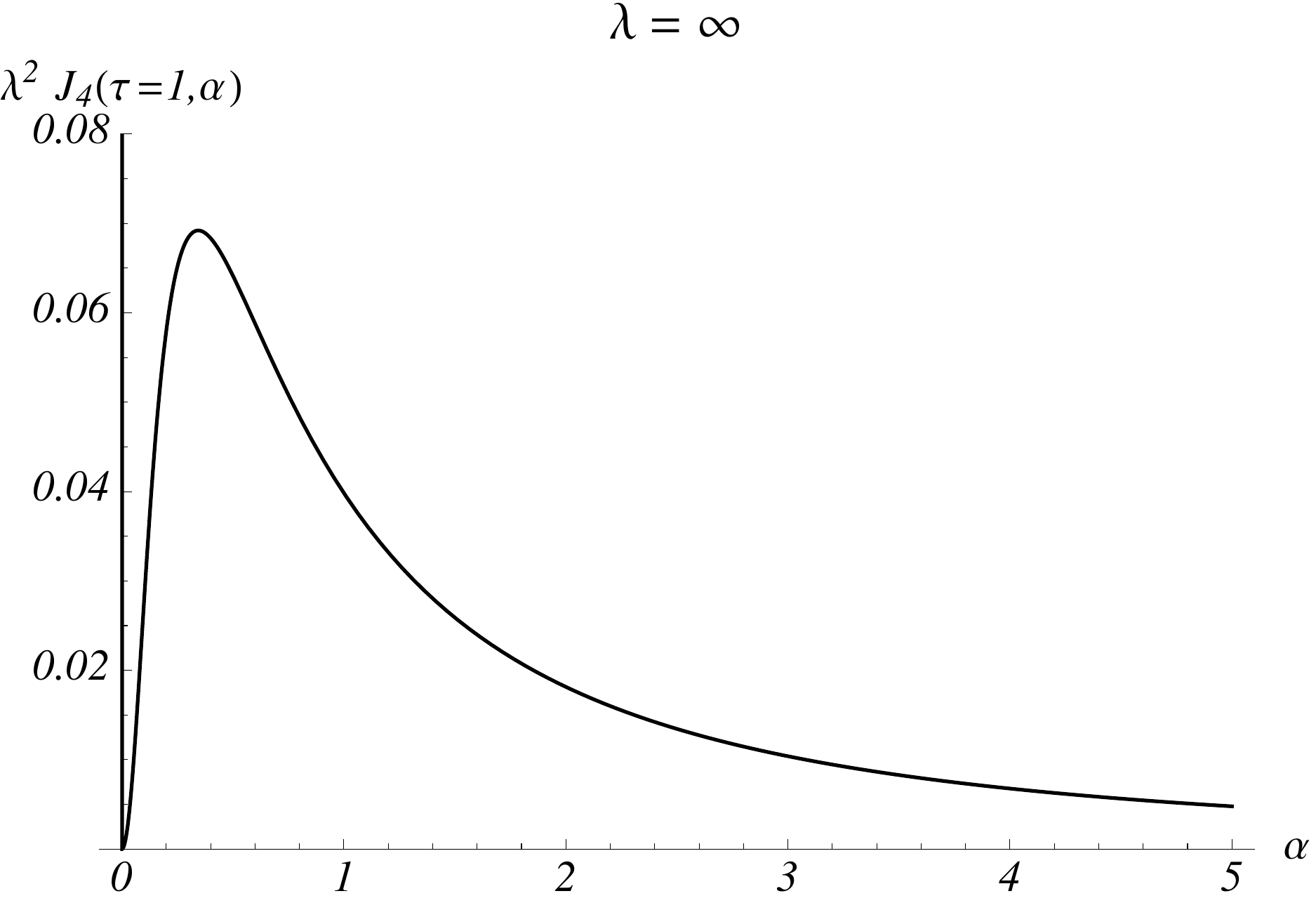}
\caption{Plot of $\lim_{\lambda\to\infty}\lambda^{2}J_{4}$ at $\tau=1$}
\label{fig:J4}
\end{center}
\end{figure}

Inspection of the expansion of $\mc J_{4}$ shows that $\lambda^{2}\,\mc J_{4}$ also admits a smooth non trivial limit
for $\lambda\to\infty$. This is closely related to the partition function. Indeed, the holonomy equation 
\be
\mbox{Tr}(\omega^{2}) = \mbox{Tr}(\omega_{\rm BTZ}^{2}),
\ee
has the remarkable feature of involving only $\mc L$, $\mc W$ and $\mc J_{4}$. It reads
\be
\label{eq:trace2}
144\,k^{2}\,\alpha^{2}\,(\lambda^{2}-9)\, \mc J_{4}-1792\,\pi^{2}\,\alpha^{2}\,\mc L^{2}-504\,\pi\,\alpha\,\tau\,k\,\mc W
-168\,\pi\,\tau^{2}\,k\,\mc L-21\,k^{2}=0.
\ee
Using (\ref{eq:Ldef}) and the analogous
 \be
 \label{eq:Wdef}
 \partial_{\alpha}\log Z=4\pi^{2}i\,\mc W,
 \ee
 we obtain after a short calculation
 \ba
\lambda^{2}\, \mc J_{4} &\stackrel{\lambda\to\infty}{=}& \frac{21}{3200\,\alpha^{4}}\,\bigg[
 -3\,\tau^{4}\,
\hyper{-\frac{3}{4}, \ -\frac{1}{2}, \ -\frac{1}{4}}{\frac{1}{3}, \ \frac{2}{3}}{-\frac{5120\,\alpha^{2}}{81\,\tau^{4}}}\nonumber \\
&&+40\,\alpha^{2}\,
 \hyper{\frac{1}{4}, \ \frac{1}{2}, \ \frac{3}{4}}{\frac{4}{3}, \ \frac{5}{3}}{-\frac{5120\,\alpha^{2}}{81\,\tau^{4}}}+40\alpha^{2}+3\,\tau^{4}
 \bigg].
 \ea
 Again, this allows to extract to large $\alpha/\tau^{2}$ expansion that is 
 \ba
 \lambda^{2}\, \mc J_{4} &\stackrel{\lambda\to\infty}{=}& 
 \frac{21}{80}\,\frac{1}{\alpha ^2}
 +\frac{7   \Gamma
   \left(-\frac{3}{4}\right) \Gamma \left(\frac{1}{4}\right)}{128 \sqrt{3} \sqrt[4]{5} \Gamma
   \left(\frac{13}{12}\right) \Gamma \left(\frac{17}{12}\right)}\,\frac{\tau}{\alpha^{5/2}}
   +\frac{63 \sqrt{\frac{3}{5}} 
   }{160}\,\frac{\tau^{2}}{\alpha^{3}}\nonumber \\
   &&
   -\frac{147 \,\sqrt{3}  \Gamma
   \left(-\frac{1}{4}\right)^2}{10240 \,5^{3/4} \Gamma \left(\frac{7}{12}\right) \Gamma
   \left(\frac{11}{12}\right)}\,\frac{\tau^{3}}{\alpha^{7/2}}
   +\frac{63}{3200}\,\frac{\tau^{4}}{\alpha^{4}}+\dots
 \ea
A plot of $\lambda^{2}\,\mc J_{4}$ as a function of $\alpha$ for $\tau=1$ is shown in Fig.~(\ref{fig:J4}).
We remark that, for higher charges, the situation is not so simple. The higher holonomy equations involve many
higher charges, even after taking a suitable scaling limit for $\lambda\to\infty$. 

\subsubsection{Subleading correction at large $\lambda$}

We remark that the $\lambda=\infty$ expansion of the partition function and $\mc J_{4}$ is enough to reconstruct its first
sub-leading correction. Indeed, for the partition function, the coefficient of the $1/\lambda^{2}$ term in the 
$\mc O(\alpha^{2n})$ contribution is simply $-\frac{3}{2}\,n\,(n-1)$ times the leading $\lambda=\infty$ value.
In other words, we have
\be
\log Z_{\lambda}(\tau, \alpha) = \log Z_{\infty}(\tau, \alpha) + \frac{1}{\lambda^{2}}\,F(\tau, \alpha) + \dots, 
\ee
with 
\ba
F(\tau, \alpha) &=& i\,\pi\,\bigg[
\frac{3}{2\,\tau}\,\hyper{\frac{1}{4}, \ \frac{1}{2}, \ \frac{3}{4}}{\frac{4}{3},\ \frac{5}{3}}{-\frac{5120}{81}\frac{\alpha^{2}}{\tau^{4}}}
+\frac{2\alpha^{2}}{\tau^{5}}\,\hyper{\frac{5}{4}, \ \frac{3}{2}, \ \frac{7}{4}}{\frac{7}{3},\ \frac{8}{3}}{-\frac{5120}{81}\frac{\alpha^{2}}{\tau^{4}}}\nonumber \\
&&-\frac{9\tau^{3}}{160\alpha^{2}}
\hyper{-\frac{3}{4}, \ -\frac{1}{2}, \ -\frac{1}{4}}{\frac{1}{3},\ \frac{2}{3}}{-\frac{5120}{81}\frac{\alpha^{2}}{\tau^{4}}}+\frac{9\tau^{3}}{160\alpha^{2}}
\bigg].
\ea
In the case of $\mc J_{4}$, the coefficient of the $1/\lambda^{4}$ term in the 
$\mc O(\alpha^{2n})$ contribution is instead $-\frac{1}{2}\,(n+3)\,(3n-5)$ times the leading $\mc O(1/\lambda^{2})$
  value. An explicit closed expression for these sub-leading correction can be easily derived by expanding (\ref{eq:trace2}).

\section{Comments and Conclusions}

In this paper, we have presented  the $\mc O(\alpha^{18})$
expansion of the partition function of the higher spin black hole in Chern-Simons gravity based on the
$\mk{hs}[\lambda]$ algebra. The result can be successfully compared with the CFT perturbative result at $\lambda=0, 1$.
For $\lambda=1$, we are able to provide the all-order expansion and resum it in closed form.
The extended data for generic $\lambda$ suggest that it could be 
interesting to match the corrections around the special points $\lambda=0,1$ by conformal perturbation 
theory along the lines of \cite{Gaberdiel:2013jpa}. 
A remarkable feature of our result is that the special limit $\lambda\to \infty$ can be taken and leads to a smooth non 
trivial result where all the higher spin charges are non vanishing, after a suitable scaling by a power of $\lambda$.
The analysis of this limit is particularly interesting because again the perturbative series 
of both the partition function and the spin-4 charge can be resummed in closed form, including the 
first sub-leading correction at large $\lambda$. 

The perturbative series that we obtained at the rather special values $\lambda=0,1,\infty$ are all apparently
asymptotic with zero radius of convergence at small chemical potential $\alpha$. In contrast, the radius of convergence
at large $\alpha$ seems to be quite large, possibly infinite. Actually, this is the reason why we presented in that cases
expansions in powers of $1/\sqrt\alpha$, assuming that there are no no non-perturbative effects, say $\sim \exp(-c/\alpha)$ or alike. This is an intriguing feature that should deserve better investigation. Indeed, for integer $\lambda=3, 4, \dots$, 
the partition function obeys algebraic equations, dependent on $\alpha$, and its perturbative series has a finite radius of 
convergence with the BTZ branch stopping at a certain critical value of the chemical potential, 
see for instance \cite{Ferlaino:2013vga}. We simply remark that the $\lambda=0,1$ limits are somewhat outside the 
strategy of derivation of the partition function because the truncation at integer $\lambda=N$ is associated with 
gravity with $SL(N, \mathbb R)$ symmetry. 

About the $\lambda\to \infty$ limit, we recall that in the complete Gaberdiel-Gopakumar duality, two scalars are present in the gravity side and the dual CFT is unitary for 
$0\le \lambda\le 1$. However, these scalars play no role in matching the black-hole partition function, as proved by the 
agreement between the \cite{Kraus:2011ds} and \cite{Gaberdiel:2012yb} calculations. 
The reason is that black holes in $AdS_{3}$ are universal and encode the thermodynamics of the CFT at
high temperature, which in two dimensions is determined by the chiral algebra. Thus, they are insensitive 
to the microscopic details of the CFT. In pure 
higher spin gravity, there is no problem in working with $\lambda>1$. For instance, as we remarked, 
an integer $\lambda=N\ge 3$ is perfectly meaningful.  It would be very interesting to see whether the large $\lambda$
limit on the gravity side could teach something about non-unitary CFTs like logarithmic cosets of 
WZW theories at fractional admissible level \cite{Creutzig:2013pda} where the analogous large $\lambda$ limit is also well-behaved~\footnote{We thank Thomas Creutzig for comments on this topic}.
Of course, a more cautious attitude can be that of considering 
the large $\lambda$ limit just a technical tool to construct a simpler version of the black hole partition function
where matching with CFT could be done by a possibly simplified calculation.

We conclude mentioning that a promising extension of this work concerns 
 the supersymmetric version of higher spin black holes \cite{Creutzig:2012ar,Creutzig:2011fe,Chen:2013oxa,Datta:2013qja,Hanaki:2012yf}. This could 
allow further tests of AdS/CFT minimal model duality exploiting known  superconformal partners
of the duality   where the infinite dimensional higher spin 
algebra is replaced by its supersymmetric extension 
$\mk{shs}[\lambda]$ \cite{Beccaria:2013wqa,Candu:2013uya,Candu:2012jq}.
\section*{Acknowledgments}

We thank Matthias Gaberdiel, Marc Henneaux,  Thomas Creutzig,  Wei Song, and Bin Chen for interesting important comments on the manuscript.

\appendix

\section{The $\mk{hs}[\lambda]$ algebra}

The $\mk{hs}[\lambda]$ algebra is spanned by generators $V^{s}_{m}$ labeled by a spin and a mode index with 
$s\ge 2$ and $|m|<s$. For the calculation presented in this paper we need a few details of it. In particular, 
the so-called {\em lone-star product} \cite{Pope:1989sr} is the associative product defined by 
\be
V^{s}_{m}\star V^{t}_{n} = \frac{1}{2}\sum_{u=1}^{s+t-1} g^{st}_{u}(m,n,\lambda)\,V^{s+t-u}_{m+n},
\ee
with 
\ba
g^{st}_{u}(m,n,\lambda) &=& \frac{(1/4)^{u-2}}{2(u-1)!}{}_{4}F_{3}\left(
\left.
\begin{array}{cc}
\frac{1}{2}+\lambda, \ \frac{1}{2}-\lambda, \ \frac{2-u}{2}, \ \frac{1-u}{2} \\
\frac{3}{2}-s, \ \frac{3}{2}-t, \ \frac{1}{2}+s+t-u
\end{array}\right | 1
\right)\times  \\
&& \sum_{k=0}^{u-1}(-1)^{k}\binom{u-1}{k}(s-1+m)_{u-1-k}\,(s-1-m)_{k}(t-1+n)_{k}(t-1-n)_{u-1-k},
\nonumber
\ea
where $(a)_{n} = a(a-1)\cdots(a-n+1)$. The cyclic trace is defined to be zero for all generators with the exception 
of $V_{0}^{1}$. The normalisation is 
\be
\mbox{Tr}(V^{s}_{m}V^{s}_{-m}) = \frac{24}{\lambda^{2}-1}\,g^{ss}_{2s-1}(m,-m,\lambda).
\ee

\section{Partition function at $\lambda=0,1$ from CFT}

The partition function at $\lambda=0,1$ has been computed in 
\cite{Kraus:2011ds}. At $\lambda=1$, we derive here the closed expression for its perturbative coefficients.
For $\lambda=0$, we provide all the necessary details to perform higher order expansions needed to 
match our extended gravity calculation.

\subsection{$\lambda=1$}

At $\lambda=1$, the partition function is 
\be
\log Z_{\lambda=1}(\tau,\alpha) = -\frac{3ik}{2\pi\tau}\int_{0}^{\infty}dx\,\bigg[
\log\bigg(1-e^{-x+\frac{2\,i\, a\, \alpha}{\tau^{2}}x^{2}}\bigg)
+\log\bigg(1-e^{-x-\frac{2\,i \, a\, \alpha}{\tau^{2}}x^{2}}\bigg)
\bigg],
\ee
where $a=\sqrt\frac{5}{12\pi^{2}}$. We can set for simplicity $k=\tau=1$.
In order to obtain the exact form of the small $\alpha$, it is convenient to 
use the following trick. We start from the identity
\be
 \frac{\partial}{\partial\alpha}\log\bigg[\bigg(1-\varepsilon\,e^{-x+\frac{2\,i\, a\, \alpha}{\tau^{2}}x^{2}}\bigg)
\bigg(1-\varepsilon \,e^{-x-\frac{2\,i \, a\, \alpha}{\tau^{2}}x^{2}}\bigg)\bigg] = 4\,a\,\sum_{p=1}^{\infty}e^{-p\,x}\,x^{2}\,\sin(2\,a\,p\,\alpha\,x^{2})\,\varepsilon^{p}.
\ee
Then, we set $\varepsilon=1$ and expand in $\alpha$
\be
\sum_{p=1}^{\infty}e^{-p\,x}\,x^{2}\,\sin(2\,a\,p\,\alpha\,x^{2}) = 
\sum_{p=1}^{\infty}\sum_{n=0}^{\infty}(-1)^{n}e^{-p\,x}\,x^{2}\frac{(2\,a\,p\,\alpha\,x^{2})^{2n+1}}{(2n+1)!}.
\ee
Integration over $x$ is straightforward
\be
\int_{0}^{\infty}dx\, e^{-px}\,x^{4n+4}=p^{-4n-5}\,\Gamma(4n+5).
\ee
The sum over $p$ can then be done using
\be
\sum_{p=1}^{\infty}\frac{1}{p^{2n+4}} = \zeta(2n+4).
\ee
Finally, we replace the zeta values at even positive integers by
\be
\zeta(2n) = (-1)^{n-1}2^{2n-1}\pi^{2n}\,\frac{B_{2n}}{(2n)!},
\ee
where $B_{n}$ are Bernoulli numbers. After some simplification, and integrating in $\alpha$, 
we arrive at the result (\ref{eq:exact1}).
%
%
%
%

\subsection{$\lambda=0$}

The partition function at $\lambda=0$ is 
\be
\log Z_{\lambda=0}(\tau,\alpha) = \frac{3ik}{\pi\tau}\int_{0}^{\infty}dx\,\bigg[
\log\bigg(1+e^{-x+\frac{6\,i\, b\, \alpha}{\tau^{2}}(x^{2}+\gamma)}\bigg)
+\log\bigg(1+e^{-x-\frac{6\,i\, b\, \alpha}{\tau^{2}}(x^{2}-\gamma)}\bigg)
\bigg],
\ee
where $b=\sqrt\frac{5}{144\pi^{2}}$ and $\gamma$ is determined by the condition
\be
\int_{0}^{\infty}dx\bigg[\frac{1}{e^{-x-i\,\epsilon\,(x^{2}+\gamma)}+1}
-\frac{1}{e^{-x+i\,\epsilon\,(x^{2}+\gamma)}+1}\bigg]=0,\qquad \epsilon=\frac{6\,b\,\alpha}{\tau^{2}}.
\ee
Setting again $k=\tau=1$, we can manipulate this condition in a similar way to what we did at $\lambda=1$
and do explicitly the integration in $x$. We  find the following efficient expansion 
\be
\sum_{q=0}^{\infty}\sum_{n=0}^{q+1}\frac{\left(4^n-2\right) \pi ^{2 n} B_{2 n} (-2)^{2 (n+q)} \epsilon ^{2 q+1}
   \left(\frac{1}{2}\right)_{n+q} (-2 q-1)_{n+q} \gamma ^{-n+q+1}}{(2 n)! \Gamma (2 q+2)}=0.
\ee
This is a condition that can be written in series of $\epsilon$
\be
0 = \left(-\gamma -\frac{\pi ^2}{3}\right) \epsilon +\left(\gamma ^2+2 \pi ^2 \gamma +\frac{7 \pi
   ^4}{3}\right) \epsilon ^3+\left(-2 \gamma ^3-10 \pi ^2 \gamma ^2-\frac{98 \pi ^4 \gamma }{3}-62
   \pi ^6\right) \epsilon ^5+\dots, 
   \ee 
and can be solved perturbatively to find $\gamma$. We obtain 
\ba
\gamma &=& -\frac{\pi ^2}{3}+\frac{16 \pi ^4 \epsilon ^2}{9}-\frac{448 \pi ^6 \epsilon
   ^3}{9}+\frac{1254656 \pi ^8 \epsilon ^6}{405}-\frac{406598656 \pi ^{10} \epsilon
   ^8}{1215}+\frac{67556569088 \pi ^{12} \epsilon ^{10}}{1215}\nonumber \\
   &&-\frac{28832947978240 \pi ^{14}
   \epsilon ^{12}}{2187}+\frac{77150044441083904 \pi ^{16} \epsilon
   ^{14}}{18225}\nonumber \\
   &&-\frac{2032300368640154533888 \pi ^{18} \epsilon
   ^{16}}{1148175}+\frac{386452372595731059441664 \pi ^{20} \epsilon ^{18}}{413343}+\dots.
   \ea
This expression can be plugged in the $\epsilon$ expansion of  $\log Z_{\lambda=0}$ that is (at $\tau=1$) 
\ba
\log Z_{\lambda=0} &=& \frac{3}{\pi}\sum_{q=0}^{\infty}\sum_{n=0}^{2q+1}
\frac{\epsilon^{2q+2}}{\Gamma (2 q+3)}\bigg[
2^{1-2 n} (-1)^q \Gamma (2 n+1) \binom{2 q+1}{n} \gamma ^{-n+2 q+1} \\
&&
   \left(\gamma  \left(2^{2 q+1}-4^n\right) \zeta (2 (n-q))+(n+1) (2 n+1)
   \left(4^q-2^{2 n+1}\right) \zeta (2 (n-q+1))\right)\bigg]\nonumber \\
   &=& \left(-\frac{3 \gamma ^2}{2 \pi }-\pi  \gamma -\frac{7 \pi ^3}{10}\right)
   \epsilon ^2+\left(\frac{\gamma ^3}{\pi }+3 \pi  \gamma ^2+7 \pi ^3 \gamma
   +\frac{31 \pi ^5}{3}\right) \epsilon ^4\nonumber \\
   &&+\left(-\frac{3 \gamma ^4}{2 \pi }-10
   \pi  \gamma ^3-49 \pi ^3 \gamma ^2-186 \pi ^5 \gamma -\frac{4191 \pi
   ^7}{10}\right) \epsilon ^6+\dots.\nonumber
  \ea
The final result is 
\ba
\log Z_{\lambda=0}(1,\alpha) &=& i\,\pi\,\bigg(
\frac{1}{2}-\frac{2 \alpha ^2}{3}+\frac{350 \alpha ^4}{27}-\frac{18850 \alpha ^6}{27}+\frac{5839250
   \alpha ^8}{81}-\frac{20616628750 \alpha ^{10}}{1701}\nonumber \\
   && +\frac{2199779356250 \alpha
   ^{12}}{729}-\frac{253710399631250 \alpha ^{14}}{243}+\frac{22024462099943968750 \alpha
   ^{16}}{45927}\nonumber \\
   &&-\frac{38909385795904480468750 \alpha ^{18}}{137781}
\bigg)+\dots,
\ea
in agreement with (\ref{eq:mainresult}) at $\lambda=0$.

\providecommand{\href}[2]{#2}\begingroup\raggedright\endgroup

\end{document}